\documentclass[twoside,12pt]{article}
\usepackage{amssymb, amsmath, amsfonts}
\usepackage{color}
\numberwithin{equation}{section}
\newtheorem{theorem}{Theorem}
\newtheorem{lemma}[theorem]{Lemma}
\newtheorem{definition}[theorem]{Definition}

\begin{document}

\title{Bi-presymplectic chains of co-rank one and related Liouville integrable systems}
\author{ Maciej B\l aszak$^{a}$, Metin G{\" u}rses$^b$,
  and Kostyantyn Zheltukhin$^c$\\}

\begin{titlepage}

\maketitle

\noindent
{\small $^a$Department of Physics, A. Mickiewicz University, Umultowska 85 , 61-614 Poznan, Poland,}
{\small e-mail blaszakm@amu.edu.pl}\\
{\small $^b$Department of Mathematics, Faculty of Sciences, Bilkent University, 06800 Ankara, Turkey,}
{\small e-mail gurses@fen.bilkent.edu.tr}\\
{\small $^c$Department of Mathematics, Faculty of Sciences Middle East Technical University 06531 Ankara, Turkey,}
{\small e-mail zheltukh@metu.edu.tr}\\

\begin{abstract}
Bi-presymplectic chains of one-forms of co-rank one are considered. The conditions in which such chains represent some Liouville integrable systems
and the conditions in which there exist related bi-Hamiltonian chains of vector fields are derived.
To present the construction of bi-presymplectic chains, the notion of dual Poisson-presymplectic pair is used and
the concept of d-compatibility of Poisson bivectors and d-compatibility of presymplectic forms is introduced.
It is shown that bi-presymplectic representation of related flow leads directly to the construction of separation coordinates
in purely algorithmic way. As an illustration
bi-presymplectic and bi-Hamiltonian chains in ${\mathbb R}^3$ are considered in detail.
\end{abstract}

{\it Keywords:}

\end{titlepage}

\section{Introduction}

Symplectic structures play an important role in the theory of
Hamiltonian dynamical systems. In the case of a non-degenerate
Poisson tensor the dual  symplectic formulation of the dynamic
can always be introduced via the inverse of the Poisson
tensor. On the other hand, many dynamical systems admit
Hamiltonian representation with degenerate Poisson tensor. For
such tensors the notion of dual presymplectic structures was
developed \cite{l,d,b,bm}.

The presymplectic picture is especially interesting for Liouville
integrable systems. There is a well developed bi-Hamiltonian theory of such systems, starting from the early work of Gelfand and Dorfman \cite{Ge}. Particularly interesting are these systems whose construction is based on Poisson pencils of the Kronecker type
\cite{GZ1,GZ2}, with polynomial in pencil parameter Casimir functions,
together with related separability theory (see \cite{7}, \cite{m3} and references quoted there in).
The important question is whether it is possible to formulate an independent, alternative
bi-presymplectic (bi-inverse-Hamiltonian in particular) theory of such systems
with related separability theory and what is the way the two theories are related to each other.

The presented paper develops the bi-presymplectic theory of
Liouville integrable systems and related separability theory in the case when the co-rank of
presymplectic forms is one. The whole formalism is based on the
notion of \emph{d-compatibility} of presymplectic forms and \emph{d-compatibility} of
Poisson bivectors.

Let us point out that although the case of co-rank one is very special,
nevertheless is of particular importance. Actually, the majority of physically interesting
Liouville integrable systems from classical mechanics belong to that class of problems. In particular it contains all systems
with quadratic in momenta first integrals whose configuration space is flat or of constant curvature.
So, it seems that the case of co-rank one is worth separate investigation.
On the other hand it is clear that in order to complete the new theory a generalization to higher co-rank
is necessary. In fact the work is in progress, although it is a non-trivial task as the systems with higher co-ranks
show specific properties not shown in the case of co-rank one.

Another question the reader can ask is about the relevance of the formalism presented. As we know, it is a well established
bi-Hamiltonian separability theory, so what can we gain when applying its dual bi-presymplectic (bi-inverse-Hamiltonian in particular)
counterpart. The answer is as follows. In the bi-Hamiltonian approach the existence of bi-Hamiltonian representation of a given
flow is a necessary condition of separability but not a sufficient one. In order to construct separation coordinates, a Poisson projection
of the second Hamiltonian structure onto a symplectic leaf of the first one has to be done. Unfortunately, it is fare from trivial
non-algorithmic procedure that should be considered separately from case to case. Moreover, there is no proof that it is always possible.
Contrary, once we find a bi-presymplectic representation of a flow considered, the construction of separation coordinates is a fully algorithmic procedure (in a generic case obviously), as the restriction of both presymplectic structures to any leaf of a given foliation is a simple task.
For this reason we do hope that the new formalism presented in the paper is relevant for the modern separability theory and hence interesting for the readers.

The paper is organized as follows. In section 2 we give some basic
information on Poisson tensors, presymplectic two-forms,
Hamiltonian and inverse Hamiltonian vector fields and dual
Poisson-presymplectic pairs. In sections 3 and 4 the concept of
d-compatibility of Poisson bivectors and d-compatibility of closed two-forms is
developed. Then, in section 5, the main properties of
bi-presymplectic chains of co-rank one are investigated. We present the conditions in which the bi-presymplectic chain
is related to some Liouville integrable system and the conditions in which the chain is bi-inverse-Hamiltonian.
The conditions in which Hamiltonian vector fields, constructed from a given bi-presymplectic chain,
constitute a related bi-Hamiltonian chain are also found. We also illustrate a construction of separation coordinates
once a bi-presymplectic chain is given.
In last sections 6, 7 and 8,  we investigate in details, with many explicit calculations and examples, a
special case of bi-presymplectic and bi-Hamiltonian chains in
$\mathbb R^3$.

Finally, let us remark that our treatment in this work is local.  Thus, even if it is not explicitly mentioned, we always restrict our considerations  to the domain $\Sigma$ of manifold $M$ where appropriate functions, vector fields and one-forms never vanish and respective Poisson tensors and presymplectic forms are of constant co-rank. In some examples we perform calculations in particular local chart from $\Sigma$.

\section{Preliminaries}

On a manifold $M$ a Poisson tensor  is a bivector with vanishing Schouten bracket.
A function $c:M\rightarrow\mathbb{R}$ is called the \emph{Casimir
function} of the Poisson operator $\Pi$ if $\Pi dc=0$. A linear combination
$\Pi_{\lambda}=\Pi_{1}-\lambda\Pi_{0}$ ($\lambda\in\mathbb{R}$) of two Poisson
operators $\Pi_{0}$ and $\Pi_{1}$ is called a \emph{Poisson pencil} if the
operator $\Pi_{\lambda}$ is Poisson for any value of the parameter $\lambda$.
In this case we say that $\Pi_{0}$ and $\Pi_{1}$ are \emph{compatible}. Having a Poisson tensor we can define
a Hamiltonian vector fields on $M$. A vector field $X_{F}$ related to a function $F\in C^\infty(M)$ by the relation
\begin{equation}
X_{F}=\Pi dF,
\end{equation}
is called the Hamiltonian vector field with respect to the
Poisson operator $\Pi$.

Further, a \emph{presymplectic} operator $\Omega$ on $M$ defines a
two-form that is closed, i.e. $d\Omega=0,$ degenerated in general. Moreover, the kernel of any presymplectic form is always an integrable
distribution. A vector field $X^{F}$ related to a function $F\in C^\infty(M)$ by the relation%
\begin{equation}
\Omega X^{F}=dF \label{1.2}%
\end{equation}
is called the inverse Hamiltonian vector field with respect to the
presymplectic operator $\Omega$.

\begin{definition}
A Poisson bivector $\Pi$ and a presymplectic form $\Omega$ are called compatible if $\Omega \Pi \Omega$ is a closed two-form.
\end{definition}

Any non-degenerate closed two form on M  is called a \emph{symplectic} form. The inverse of a symplectic
form is an \emph{implectic} operator, i.e. invertible Poisson tensor on $M$ and vice versa.
\begin{definition}
A pair $(\Pi,\, \Omega)$ is called dual implectic-symplectic pair
on $M$ if $\Pi$ is non-degenerate Poisson tensor, $\Omega$ is non-degenerate closed two-form and $\Omega\Pi=\Pi\Omega=I$.
\end{definition}
So, in the non-degenerate case, dual implectic-symplectic pair
is a pair of mutually inverse operators on $M$.
Moreover, the Hamiltonian and the inverse Hamiltonian representations are equivalent as for any implectic bivector
$\Pi$ there is a unique  dual symplectic form $\Omega=\Pi^{-1}$ and hence a vector field Hamiltonian with respect to
$\Pi$ is an inverse Hamiltonian with respect to $\Omega$.

Let us extend these considerations onto a degenerate case. In
order to do it let us generalize the concept of dual pair from \cite{bm}. Consider a manifold $M$ of an arbitrary dimension
$m$.

\begin{definition}
A pair of  tensor fields $(\Pi, \Omega)$ on $M$ of co-rank
$r$, where $\Pi$ is a Poisson tensor and $\Omega$ is a closed
two-form, is called a dual pair (Poisson-presymplectic pair) if
there exists $r$ one-forms $\alpha_i$ and $r$
linearly independent vector fields $Z_i$,
such that the following conditions are satisfied:\\
1. $\alpha_i(Z_i)=\delta_{ij}$, $i=1,2\dots r$.\\
2. $\ker \Pi=Sp\{\alpha_i:\, i=1,\dots r \}$.\\
3. $\ker \Omega=Sp\{Z_i:\, i=1,\dots r \} $.\\
4. The following partition of unity holds on $TM$, respectively on $T^*M$
\begin{equation} \label{3}
I=\Pi\Omega + \sum_{i=1}^r Z_i\otimes\alpha_i, \qquad I=\Omega\Pi + \sum_{i=1}^r \alpha_i\otimes Z_i.
\end{equation}
\end{definition}

Contrary to the non-degenerated case, for a given Poisson tensor
$\Pi$ the choice of its dual is not unique. Also for a given
presymplectic form $\Omega$ the choice of dual Poisson tensor is
not unique. The details are given in the next section. For the
degenerate case  the  Hamiltonian and the inverse Hamiltonian vector
fields are defined in the same way as for the non-degenerate case.
But for degenerate structures the notion of Hamiltonian and
inverse Hamiltonian vector fields do not coincide. For a
degenerate dual pair it is possible to find a Hamiltonian vector
field that is not inverse Hamiltonian and an inverse Hamiltonian
vector field that is not Hamiltonian. Actually, assume that
$(\Pi,\Omega)$ is a dual pair, $X_F=\Pi dF$ is a Hamiltonian
vector field and $dF=\Omega X^F$ is an inverse Hamiltonian
one-form, where $X^F$ is an inverse Hamiltonian vector field.
Having applied $\Omega$ to both sides of Hamiltonian vector field, $\Pi$ to
both sides of inverse Hamiltonian one-form and using the
decomposition (\ref{3}) we get
\begin{equation}
dF=\Omega(X_F)+\sum_{i=1}^r Z_i(F)\alpha_i, \qquad X_F=X^F-\sum_{i=1}^r\alpha_i(X^F)Z_i.
\end{equation}
It means that an inverse Hamiltonian vector field $X^F$ is
simultaneously a Hamiltonian vector field $X_F$, i.e. $X^F=X_F$,
if $dF$ is annihilated by $\ker(\Omega)$ and $X^F$ is annihilated
by $\ker(\Pi)$.

Finally, for a dual pair $(\Pi,\Omega)$, the following important relations hold
\begin{equation}
[Z_i,Z_j]=0,\quad L_{X_F}\Pi =0, \quad  L_{Z_i}\Pi =0, \quad
L_{X^F}\Omega =0, \quad  L_{Z_i}\Omega =0,
\end{equation}
where $L_X$ is the Lie-derivative operator in the direction of vector field $X$ and $[.
\,,.]$ is a commutator.

\section{D-compatibility for non-degenerate case}

In the following section we  introduce a notion of d-compatibility when a dual pair is implectic-symplectic one, i.e.
when it is of co-rank zero. Let $M$ be a
manifold of even dimension $m=2n$.

\begin{definition}
We say that a closed two-form $\Omega_1$ is  d-compatible with  a
symplectic form $\Omega_0$ if $\Pi_0\Omega_1\Pi_0$ is a Poisson
tensor and  $\Pi_0=\Omega_0^{-1}$ is dual to $\Omega_0$.
\end{definition}

\begin{definition}
We say that a Poisson tensor $\Pi_1$ is  d-compatible with  an
implectic tensor $\Pi_0$ if $\Omega_0\Pi_1\Omega_0$ is closed and
$\Omega_0=\Pi_0^{-1}$ is dual to $\Pi_0$.
\end{definition}

Now, the following theorem relates d-compatible Poisson structures,
of which one is implectic, and d-compatible closed two-forms,
of which one is symplectic.

\begin{lemma}
\item{(i)} Let  an implectic tensor $\Pi_0$ and a symplectic form $\Omega_0$
be a dual pair.  Let a Poisson tensor $\Pi_1$ be d-compatible with
$\Pi_0$. Then $\Omega_0$ and $\Omega_1=\Omega_0\Pi_1\Omega_0$ are
d-compatible
closed two-forms.\\
\item{(ii)} Let  an implectic tensor $\Pi_0$ and a symplectic form $\Omega_0$
be a dual pair.  Let a closed two-form $\Omega_1$ be d-compatible
with $\Omega_0$. Then $\Pi_0$ and $\Pi_1=\Pi_0\Omega_1\Pi_0$ are
d-compatible Poisson tensors.
\end{lemma}
{\bf Proof.}
We have $\Pi_0\Omega_0=\Omega_0\Pi_0=I$. \\
(i) The form $\Omega_0\Pi_1\Omega_0$ is closed since
$(\Pi_0,\,\Pi_1)$ are d-compatible. The forms
$(\Omega_0,\,\Omega_1)$ are d-compatible as the tensor
\begin{equation*}
\Pi_0\Omega_1\Pi_0=\Pi_0\Omega_0\Pi_1\Omega_0\Pi_0=\Pi_1
\end{equation*}
is a Poisson tensor.\\
(ii) The tensor $\Pi_1$ is Poisson since $(\Omega_0,\,\Omega_1)$
are d-compatible. The Poisson tensors $(\Pi_0,\,\Pi_1)$ are
d-compatible as the form
\begin{equation*}
\Omega_0\Pi_1\Omega_0=\Omega_0\Pi_0\Omega_1\Pi_0\Omega_0=\Omega_1
\end{equation*}
is closed. $\Box$

What is important, in the case considered the notions of
d-compatibility and compatibility of Poisson tensors are
equivalent. Actually, one can show (see for example \cite{m1})
that if $\Omega_0\Pi_1\Omega_0$ is closed (which means
d-compatibility of $\Pi_0=\Omega_0^{-1}$ and $\Pi_1$), then
$\Pi_0$ and $\Pi_1$ are compatible and vice versa, if $\Pi_0$ and
$\Pi_1$ are compatible, then $\Omega_0\Pi_1\Omega_0$ is closed and
hence $\Pi_0$ and $\Pi_1$ are d-compatible \cite{b}.

\section{D-compatibility for degenerate case}

Let us extend the notion of d-compatibility onto the degenerate case.

\begin{definition}
A closed two-form $\Omega_0$ is d-compatible with a closed two-form
$\Omega_1$ if there exists a Poisson tensor $\Pi_0$, dual to
$\Omega_0$, such that $\Pi_0\Omega_1\Pi_0$ is Poisson. Then we say that
$\Omega_1$ is d-compatible with $\Omega_0$ with respect to
$\Pi_0$.
\end{definition}

\begin{definition}
A Poisson tensor $\Pi_1$ is  d-compatible with  a
Poisson tensor $\Pi_0$ if there exists a presymplectic form $\Omega_0$, dual to
$\Pi_0$, such that $\Omega_0\Pi_1\Omega_0$ is closed. Then we say that
$\Pi_1$ is d-compatible with $\Pi_0$ with respect to
$\Omega_0$.
\end{definition}

In the rest of the paper we restrict our considerations to the simplest case, when
dual pair considered is of co-rank one and our manifold $\mathcal{M}$ is of odd
dimension $\dim \mathcal{M}=m=2n+1$.

As was mentioned in the previous section, a presymplectic form
dual to a given Poisson tensor is not unique. The set of all
presymplectic forms dual to $\Pi$ is parametrized by an arbitrary
differentiable function on $\mathcal{M}$. Moreover, as $\Pi$ is Poisson then
an arbitrary element of its one-dimensional kernel has the form
$\alpha=\mu dH$, where $\mu$ is an arbitrary differentiable
function on $\mathcal{M}$ and $H$ is a Casimir function of $\Pi$.

\begin{lemma}
Let $\Pi$ be a fixed Poisson tensor and $\Omega$ be a dual
presympectic form.  Assume that $\alpha=\mu dH\in\ker \Pi$,
$Z\in\ker \Omega$ and $\alpha(Z)=1$. A
presymplectic form $\Omega'$ is dual to $\Pi$ if and only if
\begin{equation}\label{o}
\Omega'=\Omega+dH\wedge dF,
\end{equation}
where $F$ is an arbitrary differentiable function on $\mathcal{M}$.
\end{lemma}
{\bf Proof.} First observe that $Z'=Z+\frac{1}{\mu}\Pi dF$ is an element of $\ker\Omega'$ and that $\mu Z'(H)=\mu
Z(H)=1$. Then,
\[
\Pi\Omega'=\Pi\Omega-\Pi dF\otimes dH=I-\mu Z\otimes dH-\Pi
dF\otimes dH=I-\mu Z'\otimes dH,
\]
so $\Omega'$ is dual to $\Pi$.\\
Let $\Omega$ and $\Omega'$ be presymplectic forms dual to $\Pi$.
Let $Z'\in\ker \Omega'$ and $\mu Z(H)=\mu Z'(H)=1$. We have
\begin{equation*}
\Pi\Omega= I-\mu Z\otimes dH.
\end{equation*}
\begin{equation}\label{1}
\Pi\Omega'= I-\mu Z'\otimes dH.
\end{equation}
Multiplying (\ref{1}) by $\Omega$ we get
\begin{equation*}
\Omega\Pi\Omega'= \Omega-\mu \Omega(Z')\otimes dH.
\end{equation*}
Then, using the partition of unity, we find
\begin{equation*}
(I-\mu dH\otimes Z)\Omega'= \Omega-\mu \Omega(Z')\otimes dH,
\end{equation*}
and
\begin{equation*}
\Omega'-\Omega= -\mu dH\otimes\Omega'(Z)-\mu \Omega(Z')\otimes dH.
\end{equation*}
Since $\Omega'-\Omega$ is closed form we have
\begin{equation*}
\mu \Omega(Z')=-\mu \Omega'(Z)=dF-Z(F)\alpha
\end{equation*}
and hence (\ref{o}).
$\Box$

We also have a freedom in the choice of a Poisson tensor dual to a
given two-form. The set of all Poisson tensors dual to $\Omega$ is
parametrized by an arbitrary vector field $K$ which is both Hamiltonian and inverse-Hamiltonian with respect to
a dual pair.

\begin{lemma}
Let $\Omega$ be a fixed presymplectic form and  $\Pi$ be a dual
Poisson tensor. Assume that $Z\in \ker \Omega$, $\alpha \in \ker \Pi$ and $\alpha(Z)=1$.   Let  $K$ be a vector field such that
\begin{equation}
K=\Pi dF, \quad dF=\Omega K \quad\Rightarrow\quad Z(F)=0,\quad K(\alpha)=0
\end{equation}
for some function $F$. Then, a Poisson tensor $\Pi'$ is dual to
$\Omega$ if and only if it has a form
\begin{equation}\label{11}
\Pi'=\Pi+Z\wedge K.
\end{equation}

\end{lemma}
{\bf Proof.} First we show that $\Pi'$ is Poisson. Indeed consider
a Schouten bracket
\begin{equation*}
[\Pi',\Pi']_S=-Z\wedge L_K\Pi+K\wedge L_Z \Pi-2K\wedge [Z,K]\wedge
Z.
\end{equation*}
Since $L_K\Pi=0$, $L_Z\Pi=0$ and $[Z,K]=0$, we have
$[\Pi',\Pi']_S=0$.
 Let $\alpha=\mu dH$,
 then observe that $\alpha' \in \ker\Pi'$ takes the form $\alpha'=\mu dH'=\mu dH +dF$. Moreover, $\mu Z(H)=\mu Z(H')=1$ and
\[
\Pi'\Omega=\Pi\Omega-Z\otimes\Omega K=I-\mu Z\otimes dH-Z\otimes
dF=I-\mu Z\otimes dH',
\]
so $\Pi'$ is dual to $\Omega$. \\
Let $\Pi$ and $\Pi'$ be Poisson tensors dual to $\Omega$. Let $\mu
dH\in\ker \Pi$, $\mu dH'\in\ker \Pi'$ and
$\mu\,Z(H)=\mu\,Z(H')=1$. Using the partition of unity we get
\begin{equation*}
\Omega\Pi=I-\mu dH\otimes Z
\end{equation*}
and
\begin{equation}\label{p}
\Omega\Pi'=I-\mu dH'\otimes Z.
\end{equation}
Multiplying equation (\ref{p}) by $\Pi$ we get
\begin{equation*}
\Pi\Omega\Pi'=\Pi-\mu (\Pi dH')\otimes Z
\end{equation*}
and
\begin{equation*}
(I-\mu Z\otimes dH)\Pi'=\Pi-\mu (\Pi dH')\otimes Z.
\end{equation*}
Transforming the above equality we find
\begin{equation*}
\Pi'=\Pi- \mu Z\otimes \Pi' dH- \mu (\Pi dH')\otimes Z.
\end{equation*}
As $\Pi'$ is skew-symmetric, we can put $-\mu \Pi' dH=\mu \Pi dH'=K$, so $K=\Pi dF$, $\Omega
K=dF$ and hence (\ref{11}).
$\Box$

\begin{theorem}
Let a Poisson tensor $\Pi_0$ and a closed two-form $\Omega_0$ form a
dual pair. Let $Y_0\in \ker \Omega_0$, $\mu\,dH_0\in \ker \Pi_0$ and $\mu Y_0(H_0)=1$.\vspace{0.4 cm}\\
(i) If $\Pi_1$ is a Poisson tensor d-compatible with $\Pi_0$ with respect to $\Omega_0$, then
forms $\Omega_0$ and
$\Omega_1 =\Omega_0\Pi_1\Omega_0$ are d-compatible.\vspace{0.4 cm}\\
(ii) If $\Omega_1$ is a closed two-form d-compatible with
$\Omega_0$ with respect to $\Pi_0$, then Poisson tensors $\Pi_0$ and $\Pi_1=
\Pi_0\Omega_1\Pi_0$ are d-compatible, provided that
\begin{equation}
\mu \Pi_0\Omega_1 Y_0=\Pi_0 dF
\end{equation}
for some function $F$.
\end{theorem}
{\bf Proof.}\\
(i) $\Omega_1$ is closed as $\Pi_1$ is d-compatible with $\Pi_0$. Then, $\Pi_0\Omega_1\Pi_0=\Pi_0\Omega_0\Pi_1\Omega_0\Pi_0$ is Poisson (as was shown in \cite{b}).\vspace{0.3 cm}\\
(ii) From the d-compatibility of $\Omega_0$ and $\Omega_1$ it follows
that $\Pi_1$ is Poisson. Then,
\[
\Omega_0\Pi_1\Omega_0=\Omega_0\Pi_0\Omega_1\Pi_0\Omega_0=(I-\mu\,
dH_0\otimes Y_0)\Omega_1(I-\mu\, Y_0\otimes dH_0)=\Omega_1+\mu\,
dH_0\wedge \Omega_1(Y_0).
\]
From the assumption $\Pi_0\Omega_1 \mu Y_0=\Pi_0 dF$ it follows that either
\begin{equation*}
 \Omega_1(\mu\,Y_0)= dF \quad \mbox{if} \quad Y_0(F)=0
\end{equation*}
or
\begin{equation*}
 \Omega_1(\mu\,Y_0)= dF-\mu\, Y_0(F)dH_0 \quad \mbox{if} \quad Y_0(F)\ne
0
\end{equation*}
In both cases $\Omega_0\Pi_1\Omega_0=\Omega_1+dH_0\wedge dF$ is
closed. $\Box$

\begin{theorem}\label{theorem2}
Let a Poisson tensor $\Pi_0$ and a closed two-form $\Omega_0$ form a
dual pair. Let $Y_0\in \ker \Omega_0$, $\mu\,dH_0\in \ker \Pi_0$ and $\mu Y_0(H_0)=1$.\vspace{0.3 cm}\\
(i) If $\Pi_1$ is a Poisson tensor d-compatible with $\Pi_0$ with respect to $\Omega_0$ and
\begin{equation}
X=\Pi_1 dH_0=\Pi_0 dH_1
\end{equation}
is a bi-Hamiltonian vector field, then $\Omega_0$ and
$\Omega_1=\Omega_0\Pi_1\Omega_0 + dH_1\wedge dH_0$
is d-compatible pair of presymplectic forms.\vspace{0.3 cm}\\
(ii) If $\Omega_1$ is a presymplectic form d-compatible with
$\Omega_0$ with respect to $\Pi_0$ and
\begin{equation}
\beta=\mu \Omega_0 Y_1=\mu\, \Omega_1 Y_0
\end{equation}
is a bi-presymplectic one-form, then $\Pi_0$ and
$\Pi_1=\Pi_0\Omega_1\Pi_0+X\wedge \mu Y_0$, are d-compatible Poisson
tensors if there exist some functions $F$ and $G$ such that
\begin{equation}\label{bb}
\mu\Pi_0 \Omega_0 Y_1=\Pi_0 dF,\quad \mu \Pi_0 \Omega_1 Y_1=\Pi_0 dG,
\end{equation}
where $X=\Pi_0\beta=\Pi_0 dF$.
\end{theorem}
{\bf Proof.}\\
(i) $\Omega_1$ is closed as $\Pi_1$ is d-compatible with $\Pi_0$. Then, $\Pi_0\Omega_1\Pi_0=\Pi_0\Omega_0\Pi_1\Omega_0\Pi_0$ is Poisson (as was shown in \cite{b}).\vspace{0.3 cm}\\
(ii) From (\ref{bb}) it follows that either
$Y_0(F)\neq 0$, $Y_0(G)\neq 0$ and
\begin{equation*}
 \mu \Omega_0 Y_1=dF-\mu Y_0(F)dH_0, \qquad \mu \Omega_1 Y_1=dG-\mu Y_0(G)dH_0,
\end{equation*}
\begin{equation*}
 \mu Y_1=X+\mu^2 Y_0(F)Y_0,
\end{equation*}
or $Y_0(F)=Y_0(G)=0$ and
\begin{equation*}
\mu Y_1=X,\quad \mu \Omega_0 Y_1=\Omega_0X=dF,  \quad \mu \Omega_1 Y_1=\Omega_1X=dG.
\end{equation*}
By previous theorem part (ii) the form
$\Omega_0\Pi_1\Omega_0=\Omega_0\Pi_0\Omega_1\Pi_0\Omega_0$ is
closed. Let us prove that $\Pi_1$ is a Poisson tensor. We show
that the Schouten bracket of $\Pi_1$ is zero. First observe that
\begin{equation*}
[\Pi_1,\Pi_1]_S=2[\Pi_0\Omega_1\Pi_0,X\wedge
\mu Y_0]_S+[X\wedge \mu Y_0,X\wedge \mu Y_0]_S,
\end{equation*}
as by previous theorem
$[\Pi_0\Omega_1\Pi_0,\Pi_0\Omega_1\Pi_0]_S=0$. Next
\begin{equation*}
[\Pi_0\Omega_1\Pi_0,X\wedge \mu Y_0]_S=\mu Y_0\wedge
\Pi_0d(\Omega_1X)\Pi_0-X\wedge \Pi_0d(\Omega_1\mu Y_0)\Pi_0
\end{equation*}
and
\begin{equation*}
[X\wedge \mu Y_0,X\wedge \mu Y_0]_S=2X\wedge \mu Y_0 \wedge[\mu Y_0,X].
\end{equation*}
In the case when $\Omega_0 X=dF$ and $\Omega_1
X=dG$ we have $[\mu Y_0,X]=-X(\mu)Y_0$ and the proof is completed. In the second
case
\begin{eqnarray*}
[\mu Y_0,X]&=&[\mu Y_0,\Pi_0\Omega_1
\mu Y_0]=L_{\mu Y_0}(\Pi_0\Omega_1)\mu Y_0=\Pi_0(L_{\mu Y_0}\Omega_1)\mu Y_0-(\Pi_0d\mu\wedge Y_0)\beta\\
&=&\Pi_0d(\Omega_1\mu Y_0)\mu Y_0+\beta(\Pi_0 d\mu)Y_0=\Pi_0(d\beta)\mu Y_0+\beta(\Pi_0 d\mu)Y_0\\
&=&-\Pi_0d(\mu Y_0(F))+\beta(\Pi_0 d\mu)Y_0.
\end{eqnarray*}
Also
\begin{equation}\nonumber
\mu \,\Omega_1 Y_1=\Omega_1 X+\mu Y_0(F)\beta,
\end{equation}
hence
\begin{equation}\nonumber
\Omega_1 X=dG- \mu Y_0(F)dF+[\mu Y_0(F)]^2dH_0 -\mu Y_0(G)dH_0.
\end{equation}
So,
\begin{equation}\nonumber
\Pi_0 d(\Omega_1 X)\Pi_0=-\Pi_0d(\mu Y_0(F))\wedge X.
\end{equation}
Finally
\[
\Pi_0d(\Omega_1\mu Y_0)\Pi_0=\Pi_0d\beta \Pi_0=0
\]
and the proof is completed. $\Box$

\section{Bi-presymplectic chains}
Now we are ready to present the main result of the paper.
\begin{theorem}\label{theorem3}
Assume that on $\mathcal{M}$ we have a bi-presymplectic chain of one-forms
\begin{equation}\label{beta}
\beta_i=\mu \Omega_0 Y_i=\mu \Omega_1 Y_{i-1},\quad i=1,2, \dots,
n
\end{equation}
with d-compatible pair $(\Omega_0,\Omega_1)$ with respect to some
$\Pi_0$, which starts with a kernel vector field $Y_0$ of
$\Omega_0$ and terminates with a kernel
vector field $Y_n$ of $\Omega_1$, where $\mu$ is an arbitrary function.  Then \vspace{0.4 cm}\\
(i)
\begin{equation}\label{t3e1}
\Omega_0(Y_i,Y_j)=\Omega_1(Y_i,Y_j)=0,\quad i=1,2, \dots, n.\vspace{0.4 cm}
\end{equation}
Moreover, let us assume that
\begin{equation}\label{x}
\Pi_0\beta_i=X_i=\Pi_0dH_i,\quad i=1,2, \dots, n
\end{equation}
which implies
\begin{equation}
\begin{array}{l}
\beta_i=dH_i- \mu Y_0(H_i)dH_0,\\
\mu Y_i=X_i+\mu^2 Y_i(H_0)Y_0,
\end{array}
\end{equation}
where $\Pi_0dH_0=0$. Then,\vspace{0.4 cm}\\
(ii)
\begin{equation}
\Pi_0(dH_i,dH_j)=0\quad [X_i,X_j]=0
\end{equation}
and equation (\ref{beta}) defines a Liouville integrable system.\vspace{0.4 cm}\\
Additionally, if $Y_i(H_0)=Y_0(H_i),$ then\vspace{0.4 cm}\\
(iii) Hamiltonian vector fields $X_i$ (\ref{x}) form a
bi-Hamiltonian chain
\begin{equation}
X_i=\Pi_0dH_i=\Pi_1dH_{i-1},\quad i=1,2, \dots, n
\end{equation}
where $\Pi_1= \Pi_0\Omega_1\Pi_0+X_1\wedge \mu Y_0$. The chain starts
with $H_0$, a Casimir of $\Pi_0$, and terminates with $H_n$, a
Casimir of $\Pi_1$.
\end{theorem}
{\bf Proof.}\\
(i) From (\ref{beta}) we have
\begin{equation*}
\begin{array}{l}
\Omega_0(Y_i,Y_j)=\Omega_0(Y_{i-1},Y_{j+1})\\
\Omega_1(Y_i,Y_j)=\Omega_0(Y_{i+1},Y_{j-1}).
\end{array}
\end{equation*}
Then (\ref{t3e1}) follows from
\begin{equation}\nonumber
\Omega_0(Y_i,Y_0)=0 \quad \Omega_1(Y_i,Y_n)=0.
\end{equation}
(ii) From properties of dual pair $(\Pi_0,\Omega_0)$, if
$X_i=\Pi_0dH_i$ then
\begin{equation}\nonumber
\Pi_0(dH_i,dH_j)=\Omega_0(X_i,X_j).
\end{equation}
On the other hand  as $X_i=\mu Y_i-\alpha_iY_0$ it follows that
\begin{equation}\nonumber
\Omega_0(X_i,X_j)=\Omega_0(Y_i,Y_j).
\end{equation}
(iii) We have
\begin{equation}\nonumber
\begin{array}{r}
X_i=\Pi_0dH_i= \mu \Pi_0\Omega_1Y_{i-1}= \Pi_0\Omega_1(X_{i-1}+ \mu^2 Y_0(H_{i-1})Y_0)=\\
\Pi_0\Omega_1\Pi_0dH_{i-1}+\mu Y_0(H_{i-1})X_1=\\
(\Pi_0\Omega_1\Pi_0+ X_1\wedge \mu Y_0)dH_{i-1}=\Pi_1dH_{i-1}.
\end{array}
\end{equation}
From the Theorem (\ref{theorem2}) we know that $\Pi_1$ is a
Poisson tensor d-compatible with $\Pi_0$. We have
\begin{equation}\nonumber
\begin{array}{r}
\Pi_1dH_n=(\Pi_0\Omega_1\Pi_0+ X_1\wedge \mu Y_0)dH_n=
\Pi_0\Omega_1X_n+\mu Y_0(H_n)X_1=\\
\mu\,\Pi_0\Omega_1(Y_n-\mu\,Y_0(H_n)Y_0)+\mu\,Y_0(H_n)X_1=-\mu\,Y_0(H_n)X_1+\mu\,Y_0(H_n)X_1=0
\end{array}
\end{equation}
$\Box$

A simple example of bi-presymplectic chain and its equivalent
bi-Hamiltonian representation was given in \cite{b} where the
extended Henon-Heiles system on $\mathbb{R}^5$ was considered.
Actually it is the system with Hamiltonians
\[
H_1=\frac{1}{2}p_1^2+\frac{1}{2}p_2^2+q_1^3+\frac{1}{2}q_1q_2^2-cq_1,
\]
\begin{equation}\label{HH1}
H_2=\frac{1}{2}q_2p_1p_2-\frac{1}{2}q_1p_2^2+\frac{1}{16}q_2^4+\frac{1}{4}q_1^2q_2^2-\frac{1}{4}cq_{2}^{2},
\end{equation}
where $(q,p)$ are canonical coordinates and $c$ is a Casimir
coordinate. We will come back to this example in the end
of this section.

Note that the Theorem \ref{theorem3} holds in an important special case
when (\ref{beta}) is \emph{bi-inverse-Hamiltonian}, i.e. $\beta_i=dH_i$,
$Y_0(H_i)=0,\,i=1,...,n$. Obviously it does not have a
bi-Hamiltonian counterpart until $\gamma_i\equiv Y_i(H_0)\neq 0$, but has
equivalent quasi-bi-Hamiltonian representation on $2n$ dimensional
manifold $M$. Indeed, as $\beta_i=dH_i$ then
\[
\Pi_0dH_i=\Pi_0\Omega_1\mu Y_{i-1}=\Pi_0\Omega_1(X_{i-1}+\gamma_i\mu^2 Y_0)=\Pi_0\Omega_1\Pi_0dH_{i-1}+\gamma_i\Pi_0dH_1.
\]
Notice that both Poisson structures $\Pi_0$ and
$\Pi_0\Omega_1\Pi_0$ share the same Casimir $H_0$ and all
Hamiltonians $H_i$ are independent of the Casimir coordinate $H_0=c$, so
the quasi-bi-Hamiltonian dynamics can be restricted immediately to
any common leaf $M$ of dimension $2n$
\begin{equation} \label{qh}
\pi_0 dH_i=\pi_1 dH_{i-1}+\gamma_i\pi_0 dH_1,\qquad i=1,...,n,
\end{equation}
where
\[
\pi_0=\Pi_0|_M,\quad \pi_1=(\Pi_0\Omega_1\Pi_0)|_M
\]
are restrictions of respective Poisson structures to $M$. Hence we deal with a
St{\"a}ckel system whose separation coordinates are eigenvalues of
the recursion operator $N=\pi_1\pi_0^{-1}$ \cite{mag}, provided that $N$ has
$n$ distinct and functionally independent eigenvalues at any point of $M$, i.e. we are in a generic case.

The advantage of bi-inverse-Hamiltonian representation when compared to bi-Hamiltonian ones is
that the existence of the first guarantees that the related Liouville integrable system is separable
and the construction of separation coordinates
is purely algorithmic (in a generic case), while the bi-Hamiltonian representation does not guarantee
the existence of quasi-bi-Hamiltonian representation and hence separability of related system. Moreover,
the projection of the second Poisson structure onto the symplectic foliation of the first one, in order to
construct a quasi-bi-Hamiltonian representation, is
far from being a trivial non-algorithmic procedure.

Let us illustrate the case on the example of the Henon-Heiles
system on $\mathbb{R}^4$ given by two constants of motion
\begin{equation} \label{HH2}
H_1=\frac{1}{2}p_1^2+\frac{1}{2}p_2^2+q_1^3+\frac{1}{2}q_1q_2^2,\,\,\,
H_2=\frac{1}{2}q_2p_1p_2-\frac{1}{2}q_1p_2^2+\frac{1}{16}q_2^4+\frac{1}{4}q_1^2q_2^2.
\end{equation}
On $\mathbb{R}^5$ differentials $dH_1$ and $dH_2$ have bi-inverse-Hamiltonian representation of the form
\[%
\begin{array}
[c]{l}%
\Omega_{0}Y_{0}=0\\
\Omega_{0}Y_{1}=dH_1=\Omega_{1}Y_{0}\\
\Omega_{0}Y_{2}=dH_{2}=\Omega_{1}Y_{1}\\
\qquad\ \ \ \ \ \ \,\,~0=\Omega_{1}Y_{2}\,\,
\end{array}
\]
where $\mu =1$, vector fields $Y_{i}$ are
\begin{align*}
Y_{0}  & =(0,0,0,0,1)^{T}\\
Y_{1}  &  =X_{1}+Y_1(H_{0})Y_0=(p_{1},p_{2},-3q_{1}^{2}-\frac{1}{2}q_{2}%
^{2},-q_{1}q_{2},-q_{1})^{T}\\
Y_{2}  &  =X_{2}+Y_2(H_{0})Y_0=(\frac{1}{2}q_{2}p_{2},\frac{1}{2}q_{2}p_{1}%
-q_{1}p_{1},\frac{1}{2}p_{2}^{2}-\frac{1}{2}q_{1}q_{2}^{2},\\
&  -\frac{1}{2}p_{1}p_{2}-\frac{1}{4}q_{2}^{3}-\frac{1}{2}q_{1}^{2}%
q_{2},-\frac{1}{4}q_{2}^{2})^{T}
\end{align*}
and presymplectic forms
\[
\Omega_{0}=\left(
\begin{array}
[c]{ccccc}%
0 & 0 & -1 & 0 & 0\\
0 & 0 & 0 & -1 & 0\\
1 & 0 & 0 & 0 & 0\\
0 & 1 & 0 & 0 & 0\\
0 & 0 & 0 & 0 & 0
\end{array}
\right),
\]
\begin{align*}
\Omega_{1}
&  =\left(
\begin{array}
[c]{ccccc}%
0 & -\frac{1}{2}p_{2} & -q_{1} & -\frac{1}{2}q_{2} & 3q_{1}^{2}+\frac{1}%
{2}q_{2}^{2}\\
\frac{1}{2}p_{2} & 0 & -\frac{1}{2}q_{2} & 0 & q_{1}q_{2}\\
q_{1} & \frac{1}{2}q_{2} & 0 & 0 & p_{1}\\
\frac{1}{2}q_{2} & 0 & 0 & 0 & p_{2}\\
-3q_{1}^{2}-\frac{1}{2}q_{2}^{2} & -q_{1}q_{2} & -p_{1} & -p_{2} & 0
\end{array}
\right).
\end{align*}
are d-compatible with respect to the canonical Poisson tensor dual to $\Omega_0$ one.
The chain starts with a kernel vector field $Y_{0}$ of $\Omega_{0}$ and
terminates with a kernel vector field $Y_{2}$ of $\Omega_{1}.$ On $\mathbb{R}^4$ we have
\[
\omega_{0}=\Omega_0|_{\mathbb{R}^4}=\left(
\begin{array}
[c]{cccc}%
0 & 0 & -1 & 0\\
0 & 0 & 0 & -1\\
1 & 0 & 0 & 0\\
0 & 1 & 0 & 0
\end{array}
\right),
\]
\[
\omega_{1}
 =\Omega_1|_{\mathbb{R}^4}\left(
\begin{array}
[c]{cccc}%
0 & -\frac{1}{2}p_{2} & -q_{1} & -\frac{1}{2}q_{2}\\
\frac{1}{2}p_{2} & 0 & -\frac{1}{2}q_{2} & 0\\
q_{1} & \frac{1}{2}q_{2} & 0 & 0\\
\frac{1}{2}q_{2} & 0 & 0 & 0
\end{array}
\right)
\]
and the quasi-bi-Hamiltonian representation takes the form (\ref{qh}), where
\begin{eqnarray*}
\pi_{0}&=&\Pi_0|_{\mathbb{R}^4}=\left(
\begin{array}
[c]{cccc}%
0 & 0 & 1 & 0\\
0 & 0 & 0 & 1\\
-1 & 0 & 0 & 0\\
0 & -1 & 0 & 0
\end{array}
\right)=\omega_0^{-1}, \\
\pi_{1}&=&\Pi_0 \Omega_1 \Pi_0|_{\mathbb{R}^4}= \left(
\begin{array}
[c]{cccc}%
0 & 0 & q_{1} & \frac{1}{2}q_{2}\\
0 & 0 & \frac{1}{2}q_{2} & 0\\
-q_{1} & -\frac{1}{2}q_{2} & 0 & \frac{1}{2}p_{2}\\
-\frac{1}{2}q_{2} & 0 & -\frac{1}{2}p_{2} & 0
\end{array}
\right)=\pi_0\omega_1\pi_0,
\end{eqnarray*}
$\gamma_1=-q_1$ and $\gamma_2=-\frac{1}{4}q_2^2$. Separation coordinates $(\lambda_1,\lambda_2)$, which are eigenvalues
of the recursion operator $N=\pi_1\pi_0^{-1}=\omega_0^{-1}\omega_1$, are related to $(q_1,q_2)$ coordinates by the following
point transformation
\[
q_1=\lambda_1+\lambda_2, \quad \frac{1}{4}q_2^2=-\lambda_1\lambda_2.
\]
Obviously, Hamiltonians (\ref{HH2}) do not form a related bi-Hamiltonian chain contrary to Hamiltonians (\ref{HH1}).
\section{Poisson and presymplectic structures in $\mathbb R^3$}

In this section we consider the Poisson and presymplectic
structures in  $\mathbb R^3$. In this case we have a convenient
description of the Poisson tensors and presymplectic forms and can
obtain simple conditions for compatibility. In $\mathbb R^3$ all
Poisson tensors are described by the following theorem \cite{agz}.

\begin{theorem}
Any Poisson tensor $\Pi$  in $\mathbb R^3$, except at some irregular points, has the form
\begin{equation}\label{poyson}
\Pi^{ij}=\mu\epsilon^{ijk}\partial_k H.
\end{equation}
Here $\mu$ and $H$ are some differentiable functions in $\mathbb R^3$ and $\epsilon^{ijk}$ is a Levi-Civita symbol.
\end{theorem}
\noindent Note that for the above Poisson tensor  we have $\Pi
dH=0$ that is the kernel of $\Pi$ is spanned by the form $dH$. To have
consistency we chose the function $\mu$ in (\ref{poyson}) the same
as the one used in (\ref{beta}). The compatible Poisson tensors in
$\mathbb R^3$ are characterize by the following theorem
\cite{agz}.

\begin{theorem}
Let a Poisson tensors $\Pi_0$ and $\Pi_1$ be given by
$(\Pi_0)^{ij}=\mu_0\epsilon^{ijk}\partial_k H_0$ and  $(\Pi_1)^{ij}=\mu_1\epsilon^{ijk}\partial_k H_1$,
where $\mu_0$, $\mu_1$ and $H_0$, $H_1$ are some differentiable functions. Then $\Pi_0$ and $\Pi_1$ are
compatible if and only if there exist a differentiable function $\Phi(H_0, H_1)$ such that
\begin{equation}
\mu_1=\mu_0\frac{\partial_{H_1}\Phi}{\partial_{H_0}\Phi}
\end{equation}
provided that $\partial_{H_1}\Phi=\partial \Phi/\partial H_1\ne 0$ and $\partial_{H_0}\Phi=\partial \Phi/\partial H_0\ne 0.$
\end{theorem}

For example, from the above theorem it follows that a Poisson tensor
$\Pi_0$, given by $\mu$ and a function $H_0$, and a Poisson tensor
$\Pi_1$, given by $-\mu$ and a function $H_1$, are compatible. One
should take $\Phi=H_0-H_1$. The presymplectic forms in $\mathbb
R^3$ are described by the following lemma.

\begin{lemma}\label{form}
Any closed two-form  $\Omega$  in $\mathbb R^3$ has the form
\begin{equation}\label{Th_Omega_eq1}
\Omega_{ij}=\epsilon_{ijk}Y^k,
\end{equation}
where $Y=(Y^1, Y^2, Y^3)^T$ is a divergence free  vector
\begin{equation}
\nabla\cdot Y=\partial_i Y^i=0.
\end{equation}
\end{lemma}

\vspace{0.3cm}

Note that for the above presymplectic form we have $\Omega Y=0$,
that is the kernel of $\Omega$ is spanned by the vector $Y$. Next let
us consider a dual pair.

\begin{lemma}\label{form-dual-tensor}
Consider a Poisson tensor $\Pi$,
$\Pi^{ij}=\mu\epsilon^{ijk}\partial_k H$, and a presymplectic
form $\Omega$,  $\Omega_{ij}=\epsilon_{ijk}Y^k$. Then $(\Pi,\,
\Omega)$ is a dual pair if and only if
\begin{equation}\label{l2e1}
\mu Y(H)=\mu Y^i\partial_i H=1.
\end{equation}
\end{lemma}
{\bf Proof.}
The form $\Omega$ is dual to the Poisson tensor $\Pi$ if the
following partition of the unit operator holds
\begin{equation}\nonumber
I=\Pi\Omega+\mu Y\otimes dH.
\end{equation}
The above equality  is equivalent to  (\ref{l2e1}).
$\Box$

We have a simple condition for compatibility of a Poisson tensor
and a presymplectic form.

\begin{lemma}\label{PiO}
The Poisson tensors $\Pi$, given by
$(\Pi)^{ij}=\mu\epsilon^{ijk}\partial_k H$, and the presymplectic
form $\Omega$,  given by  $(\Omega)_{ij}=\epsilon_{ijk}Y^k$, are
compatible if
\begin{equation}\label{PiO_eqn1}
Y(\mu[Y(H)])=Y^i\partial_i \,(\mu Y(H))=0.
\end{equation}
\end{lemma}
{\bf Proof.} We have
\begin{equation}\nonumber
\Omega\Pi\Omega=\mu Y(H)\Omega.
\end{equation}
The above form is given in terms of a vector $Y(H)Y$. It is closed if
\begin{equation}\nonumber
\nabla \cdot \left(\mu Y(H)Y \right)\equiv Y(\mu Y(H))=0.
\end{equation}
Since $\nabla \cdot Y=0$ the above equation is equivalent to (\ref{PiO_eqn1}). $\Box$

As a corollary of the previous lemma we have the condition for the
d-compatibility of two Poisson tensors.

\begin{lemma}\label {lemma3}
Consider a dual pair $(\Pi_0, \Omega_0)$ where the Poisson tensor
$\Pi_0$ is given by $(\Pi_0)^{ij}=\mu\epsilon^{ijk}\partial_k H_0$
and the presymplectic form $\Omega_0$ is given by
$(\Omega_0)_{ij}=\epsilon_{ijk}Y_0^k$. Then the Poisson tensor
$\Pi_1$, $(\Pi_1)^{ij}=-\mu\epsilon^{ijk}\partial_k H_1$,  is
d-compatible with the Poisson tensor $\Pi_0$  if
\begin{equation}\label{l3e1}
Y_0(\mu Y_0(H_1))=0.
\end{equation}
\end{lemma}

The condition for d-compatibility of two presimplectic forms in
$\mathbb R^3$ is given in the following lemma.

\begin{lemma}\label{d-comp-forms}
Consider a dual pair $(\Pi_0, \Omega_0)$ where the Poisson tensor
$\Pi_0$ is given by $(\Pi_0)^{ij}=\mu \epsilon^{ijk}\partial_k
H_0$ and the presymplectic form $\Omega_0$ is given by
$(\Omega_0)_{ij}=\epsilon_{ijk}Y_0^k$. Then the presymplectic
form $\Omega_1$, $(\Omega_1)_{ij}=\epsilon_{ijk}Y_1^k$, is
d-compatible with the presymplectic form $\Omega_0$  if
\begin{equation}
Y_1(H_0)\ne 0.
\end{equation}
\end{lemma}
{\bf Proof.} We have
\begin{equation}\nonumber
\Pi_0\Omega_1\Pi_0=\mu Y_1(H_0)\Pi_0
\end{equation}
Since $\Pi_0$ is a Poisson tensor, the above tensor is a Poisson tensor if $Y_1(H_0)\ne 0$. $\Box$

It turns out that in  $\mathbb R^3$ any two forms and any two Poisson tensors are d-compatible.

\begin{lemma}\label{d-copm-forms}
Let $\Omega_0$, $\Omega_1$ be two presimplectic forms in $\mathbb
R^3$, given by $(\Omega_0)_{ij}=\epsilon_{ijk}Y_0^k$ and
$(\Omega_1)_{ij}=\epsilon_{ijk}Y_1^k$. Then $\Omega_0$ and
$\Omega_1$ are d-compatible presimplectic forms.
\end{lemma}
{\bf Proof.}
Take a function $H_0$ such that $Y_0(H_0)\ne 0$ and $Y_1(H_0)\ne 0$. Define a Poisson tensor $\Pi_0$ by
$\Pi_0^{ij}=[Y_0(H_0)]^{-1}\epsilon^{ijk}\partial_k H_0$. Then by lemma \ref{form-dual-tensor}, $\Pi_0$ and
$\Omega_0$ are dual and by lemma \ref{d-comp-forms}, the forms $\Omega_0$ and $\Omega_1$ are d-compatible. $\Box$

\begin{lemma}\label{d-comp-Poisson}
Let $\Pi_0$, $\Pi_1$ be two Poisson tensors in $\mathbb R^3$,
given by $(\Pi_0)^{ij}=\mu\,\epsilon^{ijk}\partial_k H_0$ and
$(\Pi_1)^{ij}=-\mu\,\epsilon^{ijk}\partial_k H_1$. Then $\Pi_0$
and $\Pi_1$ are d-compatible Poisson tensors.
\end{lemma}
{\bf Proof.} By Darboux theorem we can find the coordinates
$(t_1,t_2,t_3)$ such that $\Pi_1$ is given by $\mu_1=1$ and
$H_1=t_1$. We can construct a closed form $\Omega_0$,
$(\Omega_0)_{ij}=\epsilon_{ijk}Y_0^k$, dual to $\Pi_0$ and such
that  $\partial_1 Y_0^1=0$. Then
\begin{equation}\nonumber
Y_0(\mu_1Y_0(H_1))=Y_0(Y_0^1)=0,
\end{equation}
so $\Omega_0$ and $\Pi_1$ are compatible. That is $\Pi_0$ and $\Pi_1$ are d-compatible.
Such a form $\Omega_0$ can be constructed as follows.
Consider the coordinate change
\begin{equation}\nonumber
u_1=t_1,\,  u_2=t_2,\, u_3=H_0(t_1,t_2,t_3).
\end{equation}
In these coordinates $\Pi_0$ is given by some $\tilde \mu_0$ and $\tilde H_0=u_3$.
Note that if a form is given by vector $\tilde Y=(A,B,C)^t$ in the $(u_1,u_2,u_3)$ coordinates then it is given by a vector
$Y=(A\partial_3 H_0, B\partial_3 H_0, C-A\partial_1 H_0-B\partial_2 H_0)$  in the $(t_1,t_2,t_3)$ coordinates.
We construct $\Omega_0$ in the $(u_1,u_2,u_3)$ coordinates in terms of the vector $\tilde Y_0=(A,B,C)^t$. First we choose
$C=(\tilde \mu)^{-1}$, so  $\tilde \mu Y_0(\tilde H_0)=1$. Hence  $\Pi_0$ and $\Omega_0$ are dual.
Then we choose $A$ such that $A\partial_3 H_0$ does not depend on $t_1$ in the $(t_1,t_2,t_3)$ coordinates,
so  $\Pi_1$ and $\Omega_0$ are compatible. Then we choose $B$ such that $\partial_1 A+\partial_2 B+\partial_3 C=0$,
so $\Omega_0$ is closed. $\Box$

\section{Bi-presymplectic chains in $\mathbb R^3$}

Consider closed two-forms $\Omega_0$ and $\Omega_1$ in some open domain of $\mathbb R^3$, given in terms of vectors
$Y_0$ and $Y_1$ by
\begin{equation}\nonumber
\Omega_{0,ij}=\epsilon_{ijk}Y_0^k \quad \mbox{where} \quad
\partial_kY_0^k=0, ~~i,j=1,2,3
\end{equation}
and
\begin{equation}\nonumber
\Omega_{1,ij}=\epsilon_{ijk}Y_1^k \quad \mbox{where} \quad
\partial_kY_1^k=0, ~~i,j=1,2,3.
\end{equation}

By lemma \ref{d-copm-forms} there exists  a Poisson tensor $\Pi_0$
such that $\Pi_0$ and $\Omega_0$ are dual and $\Omega_0$ and
$\Omega_1$ are d-compatible with respect to $\Pi_0$. We can choose a
function $H_0$ such that $\mu \,Y_0(H_0)=1$ and $Y_1(H_0)\ne 0$,
so $\Pi_0^{ij}=\mu\, \epsilon^{ijk}\partial_k H_0$. It is easy to
see that in $\mathbb R^3$ any two presymplectic forms $\Omega_0$ and $\Omega_1$ give a
bi-presymplectic chain
\begin{equation}\label{omegachain}
\begin{array}{l}
\quad \Omega_0 Y_0=0\\
\,\, \mu\,\Omega_0Y_1=\beta=\mu\, \Omega_1 Y_0\\
\,\,\,\, \qquad \qquad 0=\Omega_1Y_1.
\end{array}
\end{equation}
Then, we can consider a vector field $X$
\begin{equation}
X=\Pi_0 \beta.
\end{equation}

To construct bi-Hamiltonian representation of the above chain  we use theorem \ref{theorem3}.
Let the chain (\ref{omegachain}) be such that
\begin{equation}\label{existence_cond}
\Pi_0\beta=X=\Pi_0dH_1
\end{equation}
and hence
\begin{equation}\label{vectorbeta}
\beta=dH_1-\mu Y_0(H_1)dH_0.
\end{equation}
Then, by theorem \ref{theorem3} (ii) the vector field $X$ defines a
Liouville integrable system.

Let us obtain some relations that we will need later. Combining
(\ref{omegachain}) and (\ref{vectorbeta}) we have
\begin{equation}\label{eqn8} \nonumber
\mu\,\epsilon_{ijk}Y_0^kY_1^j=H_{1,i}-\mu\,Y_0(H_1)H_{0,i},~~i=1,2,3
\end{equation}
that gives
\begin{equation}\label{eqn9}\nonumber
Y_0(H_1)-\mu\, Y_0(H_1)Y_0(H_0)=0
\end{equation}
and
\begin{equation}\label{eqn10}\nonumber
Y_1(H_1)=\mu\,Y_0(H_1)Y_1(H_0).
\end{equation}
Using duality of $\Omega_0$ and $\Pi_0$ we have
\begin{equation}\label{eqn13}
\mu Y_1^n=\mu^2 Y_1(H_0)Y_0^n +X^n ,~~n=1,2,3.
\end{equation}
Note that if $Y_0(H_1)=0$ then $\beta$ is closed and $Y_1(H_1)=0$. So,
\begin{equation}
Y_0(H_1)=Y_1(H_1)=0.
\end{equation}

Following \cite{agz}  every Hamiltonian system in $\mathbb R^3$
has a bi-Hamiltonian representation. Thus the vector field
$X=\Pi_0dH_1$ can be also written as $X=\bar\Pi_1 dH_0$, where
$(\bar\Pi_1)^{ij}=-\mu\,\epsilon^{ijk}\partial_k H_1$ for
$i,j=1,2,3$.

Theorem \ref{theorem3} also gives the bi-Hamiltonian
representation of the vector field $X$. Let us show that these two
representations coincide. Let $Y_0(H_1)=Y_1(H_0)$ then by theorem
\ref{theorem3} (iii) we can define
\begin{equation}\label{pi}
\Pi_1=\Pi_0\Omega_1\Pi_0+\mu\,X\wedge Y_0
\end{equation}
that is
\begin{equation}\nonumber
\Pi_1^{ij}=-\mu^2\,Y_1(H_0)\epsilon^{ijk}\partial_kH_0 +
\mu\,(X^iY_0^j-X^jY_0^i),~~i,j=1,2,3
\end{equation}
Since $X^i=\epsilon^{ijk}\Pi_0^kH_{1,k}$, we can put
\begin{equation}\nonumber
X^iY_0^j-X^jY_0^i=\epsilon^{ijk}W_k,~~i,j=1,2,3
\end{equation}
So,
\begin{equation}\nonumber
\Pi_1^{ij}=-\mu^2\, Y_1(H_0)\epsilon^{ijk}\partial_kH_0 + \mu\,
\epsilon^{ijk}W_k=\epsilon^{ijk}(-\mu^2\,
Y_1(H_0)\partial_kH_0+\mu\, W_k),
\end{equation}
for all $i,j=1,2,3$. Since $\Pi_1$ is a Poisson tensor and $dH_1$
belongs to the kernel of $\Pi_1$ we have
\begin{equation}\label{eqn19}
-\mu^2\,Y_1(H_0)\partial_kH_0+\mu\, W_k=-\mu \partial_kH_1
\end{equation}
where $\mu$ is an arbitrary function.
For $W_k$ we have
\begin{eqnarray}
W_k&=&\epsilon^{ijk}X^iY_0^k=\mu\,
\epsilon^{ijk}\epsilon^{imn}H_{0,n}H_{1,m}Y_0^j=
\mu\,(\delta^n_i\delta^k_n-\delta^k_m\delta^n_j)H_{0,n}H_{1,m}Y_0^j \nonumber\\
&=&\mu\,Y_0(H_1)H_{0,k}-\mu\,Y_0(H_0)H_{1,k},~k=1,2,3\nonumber
\end{eqnarray}
where $H_{0,k}=\partial_{k}\, H_{0}$ and $ H_{1,k}=\partial_{k}\,
H_{1}$. Using the above equality for $W_k$ in (\ref{eqn19}) we get
\begin{equation}\nonumber
-\mu^2\, Y_1(H_0)\partial_kH_0 +
\mu^2\,Y_0(H_1)H_{0,k}-\mu\,H_{1,k}=-\mu H_{1,k},~k=1,2,3
\end{equation}
which gives
\begin{equation}\label{eqn21}
Y_1(H_0)= Y_0(H_1).
\end{equation}
Eqs.(\ref{eqn21}) and (\ref{eqn13}) are the only constraints on
$Y_{0}$ and $Y_{1}$. We conclude that any presymplectic chain
which fulfills the condition (\ref{existence_cond}) leads to a
bi-Hamiltonian chain.

As the next example shows, there exist presymplectic chains that
do not admit a dual bi-Hamiltonian representation.

\vspace{0.3cm}

\noindent {\bf Example 1.}\, Consider closed two-forms $\Omega_0$
and $\Omega_1$ in $\mathbb R^3$, given   by
\begin{equation}\nonumber
\Omega_{0}=
\left(
\begin{array}{rrr}
0&-1&0\\
1&0&0\\
0&0&0\\
\end{array}
\right),\quad
\Omega_{1}=
\left(
\begin{array}{rrr}
0& c & -b \\
-c&0&a\\
b& -a&0\\
\end{array}
\right).
\end{equation}
where $a,b$ and $c$ are the functions of $x_{1}$, $x_{2}$ and $x_{3}$.
Their kernels are spanned by vectors $Y_0=(0,0,1)^t$ and
$Y_1=(a,b,c)^t$ respectively. Since $\nabla \cdot Y_{1}=0$ then we have
\begin{equation}\label{div1}\nonumber
\partial_1 a+\partial_2 b+\partial_3 c=0.
\end{equation}
We take a Poisson tensor $\Pi_0$  in the form
\begin{equation}\label{ex1eqn1}\nonumber
\Pi_{0}=\mu\, \left(
\begin{array}{rrr}
0&H_{0,3}&-H_{0,2}\\
-H_{0,3}&0&H_{0,1}\\
H_{0,2}&-H_{0,1}&0\\
\end{array}
\right),
\end{equation}
where $\mu$ and $H_{0}$ are arbitrary functions of $x^{1},x^{2}$
and $x^{3}$. If $\mu H_{0,3}=1,$
then one can easily show that $\Pi_0$ and $\Omega_0$ are dual and
$\Omega_0$ and $\Omega_1$ are d-compatible with respect to
$\Pi_0$. The forms $\Omega_0$ and $\Omega_1$ make a presymplectic
chain
\begin{equation}\label{ex1eqn2}
\begin{array}{l}
\quad \Omega_0 Y_0=0\\
\,\, \mu\,\Omega_0Y_1=\beta=\mu\, \Omega_1 Y_0\\
\,\,\,\, \qquad \qquad 0=\Omega_1Y_1,
\end{array}
\end{equation}
where $\beta=\mu\,(b, -a, 0)^t$. Considering a vector field $X$
\begin{equation}\nonumber
X=\Pi_0 \beta=\mu\,(a, b, 0)^t .
\end{equation}
We find that an additional condition
\begin{equation}\nonumber
X=\Pi_0dH_1,
\end{equation}
gives
\begin{eqnarray}
a&=&H_{0,3}\, H_{1,2}-H_{0,2}\, H_{1,3},\label{denk1}\\
b&=& -H_{0,3}\, H_{1,1}+H_{0,1}\, H_{1,3},\label{denk2}\\
\mu\, (a\,H_{0,1}+b\,H_{0,2})&=&H_{0,1}\,H_{1,2}-H_{0,2}\,H_{1,1},
\label{denk3}
\end{eqnarray}
and from the constraint (\ref{eqn21}) we get
\begin{equation}\label{denk4}
H_{1,3}=a\,H_{0,1}+b\,H_{0,2}+c\, H_{0,3}.
\end{equation}
Using $a$ and $b$ from the equations (\ref{denk1}) and
(\ref{denk2}) respectively we show that (\ref{denk3}) is
identically satisfied.  Using $\mu H_{0,3}=1$ and the identity
(\ref{denk3}) in (\ref{denk4}) we get
\begin{equation}\label{denk5}
c=\mu H_{1,3}-H_{0,1}\,H_{1,2}+H_{0,2}\,H_{1,1}.
\end{equation}
As a summary we are left with the equations (\ref{denk1}),
(\ref{denk2}), (\ref{denk5}) for $a$,$b$, and $c$ and the duality
condition $\mu H_{0,3}$=1. When we use $a$, $b$ and $c$ in
(\ref{div1}) we obtain that
\begin{equation}\label{comp}
(\mu H_{1,3})_{,3}=0.
\end{equation}
This is nothing else but the d-compatibility condition (\ref{l3e1}),
i.e., $Y_{0}(\mu Y_{0}(H_{1}))=0$, of the Poisson tensors
$\Pi_{0}$ and $\Pi_{1}$. Eq. (\ref{comp}) means that
\begin{equation}\label{h1}
H_{1}=h_{1}(x^{1},x^{2})\,H_{0}+h_{2}(x^{1},x^{2})
\end{equation}
where $h_{1}$ and $h_{2}$ are arbitrary functions of $x^{1}$ and
$x^{2}$. Using (\ref{h1}) we get
\begin{eqnarray}
a&=& (h_{1,2}\, H_{0}+h_{2,2})\,H_{0,3},\label{denk6}\\
b&=& -(h_{1,1}\, H_{0}+h_{2,1})\, H_{0,3},\label{denk7}\\
c&=&h_{1}-(h_{1,2}\, H_{0}+h_{2,2})\,H_{0,1}+(h_{1,1}\,
H_{0}+h_{2,1})\, H_{0,2}. \label{denk8}
\end{eqnarray}
The above equations might be considered as differential equations
to determine $H_{0}$, $h_{1}$ and $h_{2}$ with no conditions on
$a$,$b$ and $c$. When we use (\ref{denk6}) and (\ref{denk7}) we
find that
\begin{equation}\label{denk9}
H_{0}=-{a h_{2,1}+b h_{2,2} \over a h_{1,1}+b h_{1,2}},~~
H_{0,3}={a h_{1,1}+b h_{1,2} \over h_{1,1}h_{2,2}-h_{1,2}h_{2,1}}
\end{equation}
These equations put a constraint on the $x^3$ dependence on the
given functions $a$,$b$ and $c$. Hence  we may have a
presymplectic structure with the conditions (\ref{denk9}) not
satisfied and thus obtain a presymplectic chain with no  dual
bi-Hamiltonian chain.

\section{Bi-Hamiltonian chains in $\mathbb R^3$}

Suppose we have two compatible Poisson structures $\Pi_0$ and
$\Pi_1$ in $\mathbb R^3$, given by
$(\Pi_0)_{ij}=\mu\,\epsilon^{ijk}\partial_k H_0 $  and
$(\Pi_1)_{ij}=-\mu\,\epsilon^{ijk}\partial_k H_1$, ($i,j=1,2,3$).
The Casimirs of the $\Pi_0$ and $\Pi_1$ are $dH_0$  and $dH_1$
respectively. Then we can consider a bi-Hamiltonian chain
\begin{equation}\label{bi-Ham-chain}
\begin{array}{c}
\Pi_0dH_0=0\,\,\,\,\,\\
\qquad \qquad \Pi_0dH_1=X=\Pi_1dH_0\\
\qquad \qquad \qquad \qquad \quad 0=\Pi_1dH_1.
\end{array}
\end{equation}
Using theorem 11 we can construct a corresponding bi-presymplectic
chain. To construct the bi-presymplectic chain we have to find a
closed form $\Omega_0$ dual to the Poisson structure $\Pi_0$ and
compatible with the Poisson structure $\Pi_1$. By lemma
\ref{d-comp-Poisson} such a form always exists. Having such a form
$\Omega_0$ the construction of the bi-presymplectic chain is
straightforward. We start with
$(\Omega_{0})_{ij}=-\epsilon_{ijk}\,Y_{0}^{k}, ~i,j=1,2,3$ where
\begin{equation}
\nabla \cdot Y_{0}=0, ~~\mu\, Y_{0}(H_{0})=1,
\end{equation}
\begin{equation}\label{ch5.2eqn1}
Y_{0}(\mu\,Y_{0}(H_{1}))=0
\end{equation}
and $\Omega_{1}$ is found from $ Y_1= \mu\,Y_1(H_0)Y_0 + {1 \over
\mu}\,X$. The equation (\ref{ch5.2eqn1}) is obtained from the
divergence free condition of $Y_1=\mu\, Y_1(H_0)Y_0+{1 \over
\mu}\, X$.

\vspace{0.3cm}

\noindent {\bf Example 2}\, Consider the Lorentz system \cite{agz}
\begin{equation}\nonumber
\begin{array}{lll}
\frac{d}{dt}x_{1}&=&\frac{1}{2}x_2\\
\frac{d}{dt}x_{2}&=&-x_1x_3 \\
\frac{d}{dt}x_{3}&=&x_1x_2\\
\end{array}
\end{equation}
It admits a bi-Hamiltonian representation (\ref{bi-Ham-chain})
with $H_0=\frac{1}{4}(x_3-x_1^2)$, $\mu=1$ and $H_1=x_2^2+x_3^2$.
The form $\Omega_0$ dual to $\Pi_0$ and compatible with $\Pi_1$ is
given by
\begin{equation}\nonumber
\Omega_{0}= -\left(
\begin{array}{rrr}
0& \gamma & -\beta \\
-\gamma&0& \alpha\\
b& -\alpha&0\\
\end{array}
\right),~~ \Pi_{0}= \left(
\begin{array}{rrr}
0& 1/4 & 0 \\
-1/4&0&-x_{1}/2\\
0& x_{1}/2&0\\
\end{array}
\right)
\end{equation}
where the vector $Y_0=(\alpha,\beta, \gamma)^t$. The conditions on
$\alpha, \beta $ and $\gamma$ are
\begin{equation}\nonumber
\nabla \cdot Y_{0}=\partial_1\alpha+\partial_2\beta+\partial_3\gamma=0,\quad Y_{0}(H_{0})=\frac{1}{4}\gamma-\frac{1}{2}x_{1}\alpha=1.
\end{equation}
One can find $\Omega_{1}$ having determined $Y_{1}$ from (\ref{eqn13})
\begin{equation}\nonumber
Y_{1}=({1 \over 2}x_{2}+2 \alpha \eta, -x_{1}\, x_{3}+2 \beta
\eta,x_{1} x_{2}+2 \gamma \eta)
\end{equation}
where $\eta={1 \over 2}\, Y_{0}(H_{1})=\beta x_{2}+\gamma x_{3}$.
We have an additional constraint on $\alpha, \beta $ and $\gamma$
coming from $\nabla \cdot Y_{1}=0$ which reads
\begin{equation}\nonumber
Y_{0}(\eta)=\alpha \partial_1\eta+\beta \partial_2\eta+\gamma\partial_3 \eta=0
\end{equation}
A simple solution for the above presymplectic structures is given
as $\alpha=-2/x_{1},~\beta=-2x_{2}/x_{1}^2, \gamma=0$.

It is also possible to start with a dual pair and construct a
second d-compatible Poisson structure with given properties. The
following example gives hints how to solve equations arising from
d-compatible Poisson structures.

\vspace{0.3cm}

\noindent {\bf Example 3}.  We take a dual pair $(\Pi_0,
\Omega_0)$ and construct  a Poisson  tensor $\Pi_1$, compatible
with a given pair, such that  $\Pi_1$ is non linear in $x_3$.

Let  $\Pi_0$ be given in canonical coordinates.
We take the form $\Omega_0$ as follows
\begin{equation}\nonumber
\Omega_0=
\left(
\begin{array}{rrr}
0 & -1 & f_1\\
1 & 0 & f_2\\
-f_1 & -f_2 & 0 \\
\end{array}
\right),~~\Pi_{0}=\left(
\begin{array}{rrr}
0 & 1 & 0\\
-1 & 0 & 0\\
0 & 0 & 0 \\
\end{array}
\right),
\end{equation}
where $f_1=\partial_1 f$ and $f_2=\partial_2 f$ for some function
$f(x_1,x_2)$. Note that $(\Omega_0)_{ij}=-\epsilon_{ijk}Y_0^i$
where $Y_0=(-f_2,f_1,1)$ and $H_{0}=x_{3}$. It is easy that
$\nabla \cdot Y_0=0$, so by lemma \ref{form} $\Omega_0$ is closed
and equality (\ref{l2e1}) holds, so by lemma
\ref{form-dual-tensor} is dual to $\Pi_0$. We construct a Poisson
tensor $\Pi_1$ compatible with $\Omega_0$. Let $\Pi_1$ be given by
$(\Pi_1)_{ij}=\epsilon_{ijk}\partial_k \chi $. Note that $\Pi_1$
is compatible with $\Pi_0$. By the lemma \ref{lemma3} $\Omega_0$
and $\Pi_1$ are compatible if equality (\ref{l3e1}) holds.
Consider
\begin{equation}\nonumber
Y_0\nabla \chi=-f_2\partial_1 \chi + f_2\partial_2 \chi+\partial_3 \chi.
\end{equation}
Let us perform the coordinate transformation
\begin{equation}\nonumber
\begin{array}{l}
\xi=\alpha(x_1,x_2,x_3)\\
\eta=\beta(x_1,x_2,x_3)\\
\zeta=\gamma(x_1,x_2,x_3).\\
\end{array}
\end{equation}
Then
\begin{equation}\nonumber
\begin{array}{l}
\partial_1\chi=\partial_\xi\chi\partial_1\alpha
+\partial_\eta\chi\partial_1\beta+\partial_\zeta\chi\partial_1\gamma\\
\partial_2\chi=\partial_\chi\partial_2\alpha
+\partial_\eta\chi\partial_2\beta+\partial_\zeta\chi\partial_2\gamma\\
\partial_3\chi=\partial_\xi\chi\partial_3\alpha
+\partial_\eta\chi\partial_3\beta+\partial_\zeta\chi\partial_3\gamma,\\
\end{array}
\end{equation}
so
\begin{equation}\nonumber
\begin{array}{l}
Y_{0} \cdot \nabla \chi=(-f_2\partial_1\alpha +
f_1\partial_2\alpha +
\partial_3\alpha)\partial_\xi\chi+
(-f_2\partial_1\beta + f_1\partial_2\beta +
\partial_3\beta)\partial_\eta\chi+
\\(-f_2\partial_1\gamma +
f_1\partial_2\gamma + \partial_3\gamma)\partial_{\zeta}\chi
\end{array}
\end{equation}
To simplify the above expression we choose $\beta,\,\gamma, \alpha$ such
that
\begin{equation}\nonumber
\begin{array}{l}
-f_2\partial_1\beta + f_1\partial_2\beta + \partial_3\beta=0 \\
-f_2\partial_1\gamma + f_1\partial_2\gamma + \partial_3\gamma=0\\
(-f_2\partial_1\alpha + f_1\partial_2\alpha + \partial_3\alpha)=1,
\end{array}
\end{equation}
 hence
\begin{equation}\nonumber
Y_{0} \cdot \nabla \chi=\partial_\xi\chi.
\end{equation}
Using the above technique we can solve $Y_{0}(H_{0})=1$ and in
particular $Y_{0}(Y_{0}(H_{1}))=0$ very easily. The equality
(\ref{l2e1}) holds if $H_0=\xi$. Then, $Y_{0}(Y_{0}(H_{1}))=H_{1,\xi \xi}=0$ and
\begin{equation}\nonumber
H_{1}=A_1(\zeta,\eta)\xi+A_2(\zeta,\eta),
\end{equation}
where $A_1$, $A_2$ are some arbitrary functions  of $\zeta, \, \eta$. As an application let
\begin{equation}\nonumber
\eta=x_1x_2,\quad \zeta=x_3-\ln x_2, \quad \xi=x_{3}
\end{equation}
and $f=x_1x_2=\eta$. Then, $H_0=x_{3}$ and
\begin{equation}\nonumber
H_1=A_1(x_3-\ln x_2, x_1x_2) x_{3}+A_2(x_3-\ln x_2,x_1x_2),
\end{equation}
where $A$ and $B$ are functions of $(x_{3}-\ln x_{2})$ and $x_{1}\,
x_{2}$.

\section*{Acknowledgement}

M.B. was partially supported by Polish MNiSW research grant no. N
N202 404933  and by the Scientific and Technological Research
Council of Turkey (TUBITAK). This work was partially supported by
the Turkish Academy of Sciences and by the Scientific and
Technical Research Council of Turkey.

\end{document}